\documentclass[journal,twocolumn]{IEEEtran}
\IEEEoverridecommandlockouts
\usepackage{amsthm}
\usepackage{tabularx}
\usepackage{arydshln}
\usepackage{multirow}
\usepackage{calc}
\usepackage{blkarray}
\theoremstyle{definition}

\newtheorem{example}{Example}
\newtheorem{remark}{Remark}

\usepackage{graphicx,cite}
\usepackage{dblfloatfix}

\usepackage{blindtext, graphicx, amsfonts,
	amssymb,multirow,epstopdf}
\usepackage{dsfont}

\newcolumntype{P}[1]{>{\centering\arraybackslash}p{#1}}

\ifCLASSINFOpdf

\else

\fi

\usepackage[cmex10]{amsmath}
\usepackage[linesnumbered,ruled]{algorithm2e}

\SetCommentSty{mycommfont}
\newcommand{\bfC}{\mathbf{C}}

\newcommand{\bfA}{\mathbf{A}}
\newcommand{\bfB}{\mathbf{B}}

\usepackage{url}

\usepackage[breaklinks]{hyperref}
\usepackage{breakurl}

\begin{document}
    \title{Erasure coding for distributed matrix multiplication for matrices with bounded entries}
\author{\IEEEauthorblockN{Li Tang, Konstantinos Konstantinidis
		 and Aditya Ramamoorthy \thanks{This work was supported in part by the National Science Foundation (NSF) under grant CCF-1718470. The authors are with the Department of Electrical and Computer Engineering, Iowa State University, Ames, IA 50010 USA (e-mail: \{litang, kostas, adityar\}@iastate.edu.  Li Tang and Konstantinos Konstantinidis contributed equally to this work and are joint first authors.}}
}
	\maketitle
\begin{abstract}
Distributed matrix multiplication is widely used in several scientific domains. It is well recognized that computation times on distributed clusters are often dominated by the slowest workers (called stragglers). Recent work has demonstrated that straggler mitigation can be viewed as a problem of designing erasure codes. For matrices $\bfA$ and $\bfB$, the technique essentially maps the computation of $\bfA^T \bfB$ into the multiplication of smaller (coded) submatrices. The stragglers are treated as erasures in this process. The computation can be completed as long as a certain number of workers (called the recovery threshold) complete their assigned tasks.

We present a novel coding strategy for this problem when the absolute values of the matrix entries are sufficiently small. We demonstrate a tradeoff between the assumed absolute value bounds on the matrix entries and the recovery threshold. At one extreme, we are optimal with respect to the recovery threshold and on the other extreme, we match the threshold of prior work. Experimental results on cloud-based clusters validate the benefits of our method.

%
%
%
\end{abstract}
\section{Introduction}
\label{sec:intro}
The multiplication of large-dimensional matrices is a key problem that is at the heart of several big data computations. For example, high-dimensional deep learning problems often require matrix-vector products at every iteration. In most of these problems the sheer size of the matrices precludes computation on a single machine. Accordingly, the computation is typically performed in a distributed fashion across several computation units (or workers). 
The overall job execution time in these systems is typically dominated by the slowest worker; this is often referred to as the ``straggler problem".

In recent years, techniques from coding theory have been efficiently utilized in mitigating the effect of stragglers. As pointed out in \cite{yu2018straggler} ({\it cf.} Appendix B in \cite{yu2018straggler}), this issue can be viewed as equivalent to coding for fault tolerance over a channel where the stragglers can be viewed as erasures. In the erasure coding context, a $(n,k)$ Reed-Solomon (RS) code allows for the recovery of all information symbols as long as any $k$ of the coded symbols are recovered (via polynomial interpolation). The innovative aspect of \cite{yu2018straggler} is in posing distributed matrix-vector and matrix-matrix multiplication in a form that is similar to an RS code. At a top level, the technique assigns the worker nodes the job of computing the product of smaller (coded) submatrices; these can be viewed as the symbols of a codeword. As long as enough coded symbols are received, decoding the required result is possible.

More specifically, the work of  \cite{yu2018straggler} considers the distributed computation of the product of two large matrices $\bfA^T$ and $\bfB$. Matrices $\bfA$ and $\bfB$ are first partitioned into $p\times m$ and $p\times n$  blocks of submatrices of equal size by the master node. Each worker is assumed to have enough memory to store the equivalent of a single submatrix of $\bfA$ and a single submatrix of $\bfB$. The master node does some basic processing on its end and sends appropriately coded submatrices to each worker. The workers multiply their stored (coded) submatrices and return the result to the master. The key result of \cite{yu2018straggler} shows that the product $\bfA^T \bfB$ can be recovered as long as {\it any} $\tau = pmn+p-1$ workers complete their computation; the value $\tau$ is called the recovery threshold of the computation.

Interestingly, similar ideas (relating matrix multiplication to polynomial interpolation) were investigated in a different context by Yagle \cite{yagle1995fast} in the mid 90's. However, the motivation for that work was fast matrix multiplication using pseudo-number theoretic transforms, rather than fault tolerance. There have been other contributions in this area \cite{duttaCG16,lee2018speeding,lee2017high,mallick2018rateless,wangLS18} as well, some of which predate \cite{yu2018straggler}.

\subsubsection*{Main Contributions} In this work, we demonstrate that as long as the entries in $\bfA$ and $\bfB$ are bounded by sufficiently small numbers, the recovery threshold ($\tau$) can be significantly reduced as compared to the approach of \cite{yu2018straggler}. Specifically, the recovery threshold in our work can be of the form $p'mn + p' -1$ where $p'$ is a divisor of $p$. Thus, we can achieve thresholds as low as $mn$ (which is optimal), depending on our assumptions on the matrix entries. 
We show that the required upper bound on the matrix entries can be traded off with the corresponding threshold in a simple manner. Finally, we present experimental results that demonstrate the superiority of our method via an Amazon Web Services (AWS) implementation.

%
%
%
%
\section{Problem Formulation}

Let $\bfA$ (size $v \times r$) and $\bfB$ (size $v \times t$) be two integer matrices\footnote{Floating point matrices with limited precision can be handled with appropriate scaling.}. We are interested in computing $\bfC \triangleq \bfA^T\bfB$ in a distributed fashion. Specifically, each worker node can store a $1/mp$ fraction of matrix $\bfA$ and a $1/np$ fraction of matrix $\bfB$. The job given to the worker node is to compute the product of the submatrices assigned to it. 
The master node waits for a sufficient number of the submatrix products to be communicated to it. It then determines the final result after further processing at its end. More precisely, matrices $\bfA$ and $\bfB$ are first block decomposed as follows:
\begin{align*}
	\bfA &= [A_{ij}], 0 \leq i < p, 0 \leq j < m, \text{~and}\\
	\bfB &= [B_{kl}], 0 \leq k < p, 0 \leq l < n,
\end{align*}
where the $A_{ij}$'s and the $B_{kl}$'s are of dimension $\frac{v}{p} \times \frac{r}{m}$ and  $\frac{v}{p} \times \frac{t}{n}$ respectively. The master node forms the polynomials
\begin{align*}
	\tilde{A}(s,z) &= \sum_{i,j} A_{ij} s^{\lambda_{ij}}z^{\rho_{ij}}, \text{~and} \\
	\tilde{B}(s,z) &= \sum_{k,l} B_{kl} s^{\gamma_{kl}}z^{\delta_{kl}}, 
\end{align*}
where $\lambda_{ij}, \rho_{ij}, \gamma_{kl}$ and $\delta_{kl}$ are suitably chosen integers.
Following this, the master node evaluates $\tilde{A}(s,z)$ and $\tilde{B}(s,z)$ at a fixed positive integer $s$ and carefully chosen points $z \in \{z_1, \dots, z_K\}$ (which can be real or complex) where $K$ is the number of worker nodes. Note that this only requires scalar multiplication and addition operations on the part of the master node. Subsequently, it sends matrices $\tilde{A}(s,z_i)$ and $\tilde{B}(s,z_i)$ to the $i$-th worker node. 

The $i$-th worker node computes the product $\tilde{A}^T(s,z_i) \tilde{B}(s,z_i)$ and sends it back to the master node. Let $1 \leq \tau \leq K$ denote the minimum number of worker nodes such that the master node can determine the required product (i.e., matrix $\bfC$) once {\it any} $\tau$ of the worker nodes have completed their assigned jobs. We call $\tau$ the \emph{recovery threshold} of the scheme. In \cite{yu2018straggler}, $\tau$ is shown to be $pmn+p-1$. 
\vspace{-0.1in}
\section{Reduced Recovery Threshold codes}
\label{sec:RedCode}
\subsection{Motivating example}
Let $m=n=p=2$ so that the following block decomposition holds
\begin{align*}
\bfA = \begin{bmatrix}
A_{00}& A_{01}\\
A_{10}& A_{11}
\end{bmatrix} \text{~and~}
\bfB = \begin{bmatrix}
B_{00}& B_{01}\\
B_{10}& B_{11}
\end{bmatrix}.
\end{align*}
We let
 \begin{align*}
  	\tilde{A}(s,z) &= A_{00}  + A_{10} s^{-1} + (A_{01} + A_{11} s^{-1})z, \text{~and}\\
 \tilde{B}(s,z) &= B_{00} + B_{10} s + (B_{01} + B_{11} s) z^2.
 \end{align*}

The product $\tilde{A}^T(s,z)\tilde{B}(s,z)$ can be verified to be
\begin{align}
&\tilde{A}^T(s,z)\tilde{B}(s,z) =  \nonumber\\
&s^{-1}(A^T_{10}B_{00}+A_{11}^TB_{00}z+A_{10}^TB_{01}z^2+A_{11}^TB_{01}z^3)\\
&+C_{00} + C_{10} z+ C_{01} z^2 + C_{11} z^3 \label{eq:useful_terms}\\
&+s(A_{00}^TB_{10} + A_{01}^TB_{10} z +A_{00}^TB_{11} z^2+A_{01}^TB_{11} z^3).
\end{align}

Evidently, the product above contains the useful terms in (\ref{eq:useful_terms}) as coefficients of $z^k$ for $k= 0, \dots, 3$. The other two lines contain terms (coefficients of $s^{-1}z^k$ and $sz^k$, $k=0, \dots, 3$) that we are not interested in; we refer to these as \emph{interference terms}. Rearranging the terms, we have
\begin{align*}
&\tilde{A}^T(s,z)\tilde{B}(s,z) = \\
&\underbrace{(* s^{-1} + C_{00}  + * s)}_{X_{00}} + \underbrace{(* s^{-1} + C_{10} + * s)}_{X_{10}} z +\\
& \underbrace{(* s^{-1} + C_{01} + * s)}_{X_{01}} z^2 + \underbrace{(* s^{-1} + C_{11} + * s)}_{X_{11}} z^3,
\end{align*}
where $*$ denotes an interference term.

As the above polynomial is of $z$-degree 3, equivalently we have presented a coding strategy where we recover superposed useful and interference terms even in the presence of $K-4$ erasures.

Now, suppose that the absolute value of each entry in $\bfC$ and of each of the interference terms is $< L$. Furthermore, assume that $s \geq 2L$. 
The $C_{ij}$'s can then be recovered by exploiting the fact that $s \ge 2L$, e.g., for non-negative matrices $\bfA$ and $\bfB$, we can simply extract the integer part of each $X_{ij}$ and compute its remainder upon division by $s$. The case of general $\bfA$ and $\bfB$ is treated in Section \ref{sec:num_trans_codes}.


To summarize, under our assumptions on the maximum absolute value of the matrix $\bfC$ and the interference matrix products, we can obtain a scheme with a threshold of $4$. In contrast, the scheme of \cite{yu2018straggler} would have a threshold of $9$. 

\begin{remark}
We emphasize that the choice of polynomials $\tilde{A}(s,z)$ and $\tilde{B}(s,z)$ are quite different in our work as compared to \cite{yu2018straggler}; this can be verified by setting $s=1$ in the expressions. In particular, our choice of polynomials deliberately creates the controlled superposition of useful and interference terms (the choice of coefficients in \cite{yu2018straggler} explicitly avoids the superposition). We unentangle the superposition by using our assumptions on the matrix entries later. To our best knowledge, this unentangling idea first appeared in the work of \cite{yagle1995fast}, though its motivations were different.
\end{remark}

\vspace{-0.1in}
\subsection{General code construction}
\label{sec:num_trans_codes}
We now present the most general form of our result. Let the block decomposed matrices $\bfA$ and $\bfB$ be of size $p \times m$ and $p \times n$ respectively. We form the polynomials $\tilde{A}(s,z)$ and $\tilde{B}(s,z)$ as follows
\begin{align*}
\tilde{A}(s,z) =& \sum_{i=0}^{m-1} z^i\sum_{u=0}^{p-1}{A_{ui}s^{-u}}, \text{~and}\\
\tilde{B}(s,z) =& \sum_{j=0}^{n-1} z^{mj}\sum_{v=0}^{p-1}{B_{vj}s^{v}}.
\end{align*}
Under this choice of polynomials $\tilde{A}(s,z)$ and $\tilde{B}(s,z)$, we have
\begin{align}
	\label{eq:multi}
	\tilde{A}^T(s,z)\tilde{B}(s,z) = \sum_{i = 0}^{m-1}\sum_{j=0}^{n-1}\sum_{u=0}^{p-1}\sum_{v=0}^{p-1}A_{ui}^TB_{vj}z^{mj+i}s^{v-u}.
\end{align}
To better understand the behavior of this sum, we divide it into the following cases.
\begin{itemize}
	\item {\it Case 1: Useful terms.} These are the terms with coefficients of the form $A_{ui}^TB_{uj}$. They are useful since $C_{ij} = \sum_{u=0}^{p-1} A_{ui}^T B_{uj}$. It is easy to check that the term $A_{ui}^TB_{uj}$ is the coefficient of $z^{mj+i}$.

	\item {\it Case 2: Interference terms.} Conversely, the terms in (\ref{eq:multi}) with coefficient $A_{ui}^TB_{vj}$, $u\neq v$ are the \emph{interference terms} and they are the coefficients of $z^{mj+i}s^{v-u}$ (for $v \neq u$). 
	
\end{itemize}

Based on the above discussion, we obtain
\begin{align}
\label{eq:final_multi}
&\tilde{A}^T(s,z)\tilde{B}(s,z) =\sum_{i=0}^{m-1}\sum_{j=0}^{n-1}z^{mj+i} \times \nonumber\\ & (\underbrace{*s^{-(p-1)}+\cdots+*s^{-1}+C_{ij}+*s+\cdots + *s^{p-1}}_{X_{ij}}),
\end{align}
where $*$ denotes an interference term. Note that (\ref{eq:final_multi}) consists of consecutive powers $z^k$ for $k = 0, \dots, mn-1$.


We choose distinct values $z_i$ for worker $i$ (real or complex). Suppose that the absolute value of each $C_{ij}$ and of each  interference term (marked with $*$) is at most $L-1$. We choose $s \ge 2L$. 
\vspace{-0.1in}
\subsection{Decoding algorithm}
\label{sec:dec_algo}


We now show that as long as at least $mn$ of the worker nodes return their computations, the master node can recover the matrix $\bfC$. 

Suppose the master node obtains the result $Y_i = \tilde{A}^T(s,z_i)\tilde{B}(s,z_i)$ from {\it any} $mn$ workers $i_1,i_2,\ldots,i_{mn}$. Then, it can recover $X_{ij}$, $i = 0, \ldots, m - 1, j = 0, \ldots, n - 1$ by solving the following equations,
\begin{align*}
\begin{bmatrix}
Y_{i_1}\\
Y_{i_2}\\
\vdots\\
Y_{i_{mn}}
\end{bmatrix}
= \begin{bmatrix}
1 & z_{i_1} & z_{i_1}^2 &\cdots &z_{i_1}^{mn-1}\\
1 & z_{i_2} & z_{i_2}^2 &\cdots &z_{i_2}^{mn-1}\\
&  & \cdots &\cdots &\\
1 & z_{i_{mn}} & z_{i_{mn}}^2 & \cdots & z_{i_{mn}}^{mn-1}\\
\end{bmatrix}
\begin{bmatrix}
X_{00}\\
X_{01}\\
\vdots\\
X_{(m-1)(n-1)}
\end{bmatrix}.
\end{align*}
The Vandermonde form of the above matrix guarantees the uniqueness of the solution. This is because the determinant of Vandermonde matrix can be expressed as $\prod_{1\le a,b\le mn} (z_{i_a}-z_{i_b})$, which is non-zero since $z_{i_j}$, $j=1,\cdots,mn$, are distinct.

Note that $X_{ij} = *s^{-(p-1)}+\cdots+*s^{-1}+C_{ij}+*s+\cdots + *s^{p-1}$. 
The master node can recover $C_{ij}$ from $X_{ij}$ as follows. We first round $X_{ij}$ to the closest integer. This allows us to recover $C_{ij}+*s+\cdots + *s^{p-1}$. This is because
\begin{align*}
|*s^{-(p-1)}+\cdots+*s^{-1}| \leq \frac{L-1}{2L - 1} < 1/2.
\end{align*}
Next, we determine  $\hat{C}_{ij} = C_{ij}+*s+\cdots + *s^{p-1}\mod s$ (we work under the convention that the modulo output always lies between $0$ and $s-1$). It is easy to see that if $\hat{C}_{ij}\le s/2$ then $C_{ij} = \hat{C}_{ij}$, otherwise $C_{ij}$ is negative and $C_{ij} = - (s - \hat{C}_{ij})$.
If $s$ is a power of $2$, the modulo operation can be performed by simple bit-shifting; this is the preferred choice.


\vspace{-0.1in}
\subsection{Discussion of precision issues}
The maximum and the minimum values (integer or floating point) that can be stored and manipulated on a computer have certain limits. 
Assuming $s = 2L$, it is easy to see that $|X_{ij}|$ is at most $(2L)^p/2$. Therefore, large values of $L$ and $p$ can potentially cause numerical issues (overflow and/or underflow).
We note here that a simple but rather conservative way to estimate the value of $L$ would be to set it equal to $v \cdot \max |\bfA| \times \max |\bfB| + 1$.

\vspace{-0.1in}
\section{Trading off precision and threshold}
The method presented in Section \ref{sec:RedCode} achieves a threshold of $mn$ while requiring that the LHS of (\ref{eq:final_multi}) remain with the range of numeric values that can be represented on the machine.
In general, the terms in (\ref{eq:final_multi}) will depend on the choice of the $z_i$'s and the values of the $|X_{ij}|$'s, e.g., choosing the $z_i$'s to be complex roots of unity will imply that our method requires $mn \times (2L)^p/2$ to be within the range of values that can be represented.

We now present a scheme that allows us to trade off the precision requirements with the recovery threshold of the scheme, i.e., we can loosen the requirement on $L$ and $p$ at the cost of an increased threshold.




Assume that $p'$ is an integer that divides $p$. We form the polynomials $\tilde{A}(s,z)$ and $\tilde{B}(s,z)$ as follows,
\begin{align*}
\tilde{A}(s,z) =& \sum_{i=0}^{m-1} \sum_{j = 0}^{p'-1}z^{j+p'i}\sum_{k=0}^{p/p'-1}A_{(k+\frac{p}{p'}j), i}s^{k}, \text{~and} \\
\tilde{B}(s,z) =& \sum_{u=0}^{n-1} \sum_{v = 0}^{p'-1}z^{mp'u+(p'-1-v)}\sum_{w=0}^{p/p'-1} {B_{(w+\frac{p}{p'}v),u}s^{-w}}.
\end{align*}
Note that in the expressions above we use $A_{i,j}$ to represent the $(i,j)$-th entry of $\bfA$ (rather than $A_{ij}$).
Next, we have
\begin{align}
\label{eq:polyABRed}
&\tilde{A}(s,z)^T \tilde{B}(s,z) = \sum_{i=0}^{m-1} \sum_{j = 0}^{p'-1}\sum_{k=0}^{p/p'-1}\sum_{u=0}^{n-1}\sum_{v = 0}^{p'-1}\sum_{w=0}^{p/p'-1}\nonumber\\&A_{(k+\frac{p}{p'}j),i}^TB_{(w+\frac{p}{p'}v),u}z^{mp'u+(p'-1-v)+j+p'i}s^{k-w}.
\end{align}
To better understand the behavior of (\ref{eq:polyABRed}), we again divide it into useful terms and interference terms.
\begin{itemize}
	\item {\it Case 1: Useful terms.} These are the terms with coefficients of the form $A_{(k+\frac{p}{p'}j),i}^TB_{(k+\frac{p}{p'}j),u}$. 
The term $A_{(k+\frac{p}{p'}j),i}^TB_{(k+\frac{p}{p'}j),u}$ is the coefficient of $z^{mp'u+p'i+p'-1}$.

	\item {\it Case 2: Interference terms.} The interference terms are associated with the terms with coefficient $A_{(k+\frac{p}{p'}j),i}^TB_{(w+\frac{p}{p'}v),u}$, $k\neq w$ and/or $j\neq v$. They can be written as
	\begin{align*}
	A_{(k+\frac{p}{p'}j),i}^TB_{(w+\frac{p}{p'}v),u}z^{mp'u+(p'-1-v)+j+p'i}s^{k-w}.
	\end{align*}
\end{itemize}
We now verify that the interference terms and useful terms are distinct. This is evident when $k \neq w$ by examining the exponent of $s$. When $k = w$ but $j \neq v$ we argue as follows. Suppose that there exist some $u_1, u_2, i_1, i_2$ such that $mp'u_1+p'i_1+p'-1=mp'u_2+p'+p'i_2-v+j-1$. Then, $mp'(u_1-u_2)+p'(i_1-i_2)=j-v$. This is impossible since $|j-v|<p'$.

Next, we discuss the degree of $\tilde{A}(s,z)^T \tilde{B}(s,z)$ in the variable $z$. In (\ref{eq:polyABRed}), the terms with maximal $z$-degree  are the terms with $u=n-1, v = 0, j = p'-1$ and $i = m - 1$. Thus, the maximal degree of $z$ in the expression is $mnp'+p'-2$. It can be verified that terms with $z$-degree from $0$ to $mnp'+p'-2$ will appear in (\ref{eq:polyABRed}) and the $z$-degree of the useful terms $C_{iu}$ are
$mp'u+p'i+p'-1$, $i = 0, \cdots, m -1$, $u = 0, \cdots, n -1$.

Likewise the $s$-degree of $\tilde{A}(s,z)^T \tilde{B}(s,z)$ varies from $-(p-1), \dots, 0, \dots, (p-1)$ with the useful terms corresponding to $s^0$. Based on the above discussion, we obtain
\begin{align*}
&\tilde{A}^T(s,z)\tilde{B}(s,z) = \sum_{k=0}^{mnp'+p'-2}  X_k z^k, \text{where}\\
&X_k =\left\{
\begin{aligned}
*s^{-(\frac{p}{p'}-1)}+\cdots+*s^{-1}+C_{ij}+*s+\cdots + *s^{\frac{p}{p'}-1},&\\
\text{if~} k = mp'j + p'i+p-1&\\
*s^{-(\frac{p}{p'}-1)}+\cdots+*s^{-1}+*+*s+\cdots + *s^{\frac{p}{p'}-1},&\\
\text{otherwise}.
\end{aligned}
\right.
\end{align*}
Evidently, the recovery threshold is $mnp'+p'-1$, which is higher than that of the construction in  Section \ref{sec:num_trans_codes}. However, let $s = 2L$, the maximum value of $|X_{ij}|$ is at most $(2L)^{p/p'}/2$ which is less than the previous construction if $p' > 1$. 

\begin{example}
	Let $m=n=2$, $p=4$ and $p'=2$ so that
	\begin{align*}
	\bfA = \begin{bmatrix}
	A_{00}& A_{01}\\
	A_{10}& A_{11}\\
	A_{20}& A_{21}\\
	A_{30}& A_{31}
	\end{bmatrix} \text{~and~}
	\bfB = \begin{bmatrix}
	B_{00}& B_{01}\\
	B_{10}& B_{11}\\
	B_{20}& B_{21}\\
	B_{30}& B_{31}
	\end{bmatrix}.
	\end{align*}
	We let
	\begin{align*}
	\tilde{A}(s,z) =& A_{00}  + A_{10} s^{-1} + (A_{20} + A_{30} s^{-1})z + \\&(A_{01} +
	A_{11}s^{-1})z^2+ (A_{21}+ A_{31}s^{-1})z^3, \text{~and}\\
	\tilde{B}(s,z) =& (B_{00} + B_{10} s)z + B_{20} + B_{30} s + \\&(B_{01} + B_{11} s)z^5 + (B_{21} + B_{31} s)z^4.
	\end{align*}
The product of the above polynomials can be verified to contain the useful terms with coefficients $z, z^3, z^5, z^7$; the others are interference terms. For this scheme the corresponding $|X_{ij}|$ can at most be $2L^2$, though the recovery threshold is 9. Applying the method of Section \ref{sec:num_trans_codes} would result in the $|X_{ij}|$ values being bounded by $8L^4$ with a threshold of $4$.
\end{example}

\begin{figure}[t]
\centering
\includegraphics[scale=0.5]{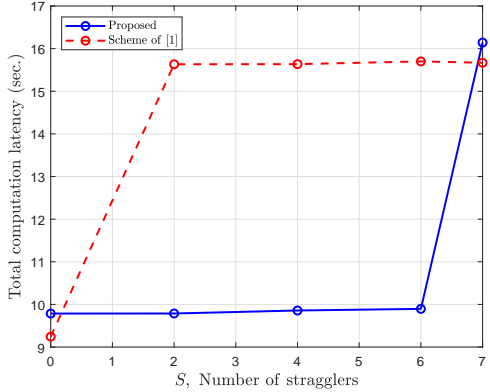}
\caption{Comparison of total computation latency by simulating up to 8 stragglers}
\label{fig:comp_latency}
\vspace{-0.2in}
\end{figure}
\vspace{-0.1in}
\section{Experimental Results and Discussion}

We ran our experiments on  AWS EC2 r3.large instances. Our code is available online\cite{commBitbuket}.  The input matrices $\bfA$ and $\bfB$ were randomly generated integer matrices of size $8000\times 8000$ with elements in the set $\{0,1,\dots,50\}$. These matrices were pre-generated (for the different straggler counts) and remained the same for all experiments. The master node was responsible for the $2 \times 2$ block decomposition of $\bfA$ and $\bfB$, computing $\tilde{A}(s,z_i)$ and $\tilde{B}(s,z_i)$ for $i=1, \dots, 10$ and sending them to the worker nodes.
The evaluation points ($z_i$'s) were chosen as 10 equally spaced reals within the interval $[-1,1]$. The stragglers were simulated by having $S$ randomly chosen machines perform their local computation twice.

We compared the performance of our method ({\it cf.} Section \ref{sec:RedCode}) with \cite{yu2018straggler}. For fairness, we chose the same evaluation points in both methods. In fact, the choice of points in their code available online \cite{nips2017Github} (which we adapted for the case when $p>1$), provides worse results than those reported here.

\textit{Computation latency} refers to the elapsed time from the point when all workers have received their inputs until enough of them finish their computations accounting for the decoding time. The decoding time for our method is slightly higher owing to the modulo $s$ operation ({\it cf.} Section \ref{sec:dec_algo}). 

It can be observed in Fig. \ref{fig:comp_latency} that for our method there is no significant change in the latency for the values of $S\in\{0,2,4,6\}$ and it remains around 9.83 seconds. When $S=7$, as expected the straggler effects start impacting our system and the latency jumps to approximately 16.14 seconds. In contrast, the performance of  \cite{yu2018straggler} deteriorates in the presence of two or more stragglers (average latency $\geq$ 15.65 seconds).

Real Vandermonde matrices are well-known to have bad condition numbers. The condition number is better when we consider complex Vandermonde matrices with entries from the unit circle \cite{gautschi1990stable}. In our method, the $|X_{ij}|$ and $|Y_{ij}|$ values can be quite large. This introduces small errors in the decoding process. Let $\hat{\bfC}$ be the decoded matrix and $\bfC\triangleq \bfA^T\bfB$ be the actual product. Our error metric is $e=\frac{||\bfC-\hat{\bfC}||_F}{||\bfC||_F}$ (subscript $F$ refers to the Frobenius norm). The results in Fig. 1, had an error $e$ of at most $10^{-7}$. We studied the effect of increasing the average value of the entries in $\mathbf{A}$ and $\mathbf{B}$ in Table 1. The error is consistently low up to a bound of $L=1000$, following which the calculation is useless owing to numerical overflow issues.  We point out that in our experiments the error $e$ was identically zero if the $z_i$'s were chosen from the unit circle. However, this requires complex multiplication, which increases the computation time.



\begin{table}[t]
\centering
\caption{Effect of bound ($L$) on the decoding error}
\label{table:error}
\begin{tabular}{|P{1cm}||P{0.5cm}|P{1.5cm}|}
\hline
Bound($L$)&s&Error\\
\hline
100&$2^{28}$&$6.31\cdot10^{-7}$\\
\hline
200&$2^{30}$&$8.87\cdot10^{-7}$\\
\hline
500&$2^{32}$&$6.40\cdot10^{-6}$\\
\hline
1000&$2^{34}$&$9.52\cdot10^{-6}$\\
\hline
2000&$2^{36}$&$1$\\
\hline
\end{tabular}
\vspace{-0.2in}
\end{table}

\vspace{-0.1in}
\bibliographystyle{IEEETran}
\bibliography{refsCL}
 \end{document}